\documentclass[aoas,preprint]{imsart}
\usepackage{bm}
\RequirePackage[OT1]{fontenc}
\RequirePackage{amsthm,amsmath}
\RequirePackage{natbib}
\setcitestyle{authoryear,open={(},close={)}}

\RequirePackage[colorlinks,citecolor=blue,urlcolor=blue]{hyperref}
\usepackage{graphicx}
 \usepackage{multirow}
 \usepackage{lscape}
\usepackage{tablefootnote}
\usepackage{ulem}
\usepackage{amssymb}
\usepackage{algorithm}% http://ctan.org/pkg/algorithm
\usepackage{algpseudocode}
\usepackage{xcolor}
\usepackage{soul}
\usepackage{amsmath}
\usepackage{enumerate}
\usepackage{natbib}
\usepackage{url} % not crucial - just u 

\usepackage{lineno}
\arxiv{arXiv:0000.0000}

\startlocaldefs
\numberwithin{equation}{section}
\theoremstyle{plain}

\endlocaldefs

\begin{document}

\begin{frontmatter}
\title{A Bayesian Time-Varying Effect Model for Behavioral mHealth Data}
\runtitle{TVEM for mHealth Data}

\begin{aug}
\author{\fnms{Matthew D.} \snm{Koslovsky}\thanksref{t1}\ead[label=e1]{mkoslovsky12@gmail.com}},

\author{\fnms{Emily T.} \snm{H\'ebert}\thanksref{t2}\ead[label=e2]{Emily-Hebert@ouhsc.edu}},

\author{\fnms{Michael S.} \snm{Businelle}\thanksref{t2}\ead[label=e3]{ Michael-Businelle@ouhsc.edu }},

\and
\author{\fnms{Marina} \snm{Vannucci}\thanksref{ t3}\ead[label=e4]{marina@rice.edu}}

\runauthor{M. Koslovsky et al.}

\affiliation{Colorado State University\thanksmark{t1}, Oklahoma Tobacco Research Center\thanksmark{t2}and Rice University\thanksmark{t3}}

\address{Department of Statistics\\ Colorado State University\\ Fort Collins, CO, USA\\
\printead{e1}\\
\phantom{E-mail:mkoslovsky12@gmail.com\ }}

\address{Oklahoma Tobacco Research Center\\
The University of Oklahoma Health Sciences Center\\
655 Research Parkway, Suite 400\\
Oklahoma City, OK 73104\\
\printead{e2}\\ \printead{e3} }

\address{Department of Statistics\\ Rice University\\ Houston, TX, USA\\
\printead{e4}}

\end{aug}

\begin{abstract}
The integration of mobile health (mHealth) devices into behavioral health research has fundamentally changed the way researchers and interventionalists are able to collect data as well as deploy and evaluate intervention strategies. In these studies, researchers often collect intensive longitudinal data (ILD) using ecological momentary assessment methods, which aim to capture psychological, emotional, and environmental factors that may relate to a behavioral outcome in near real-time. In order to investigate ILD collected in a novel, smartphone-based smoking cessation study, we propose a Bayesian variable selection approach for time-varying effect models, designed to identify dynamic relations between potential risk factors and smoking behaviors in the critical moments around a quit attempt.  We use parameter-expansion and data-augmentation techniques to efficiently explore how the underlying structure of these relations varies over time and across subjects. We achieve deeper insights into these relations by introducing nonparametric priors for regression coefficients that cluster similar effects for risk factors while simultaneously determining their inclusion. Results indicate that our approach is well-positioned to help researchers effectively evaluate, design, and deliver tailored intervention strategies in the critical moments surrounding a quit attempt. 
\end{abstract}


\end{frontmatter}

\section{Introduction}
\label{sec:intro}

\subsection{Scientific Background}

 The integration of mobile health (mHealth) devices into behavioral health research has fundamentally changed the way researchers and interventionalists are able to collect data as well as deploy and evaluate intervention strategies. Leveraging mobile and sensing technologies, just-in-time adaptive interventions (JITAI) or ecological momentary interventions are designed to provide tailored support to participants based on their mood, affect, and socio-environmental context \citep{heron2010ecological, nahum2017just}. In order to deliver theory-based interventions at critical moments, researchers collect intensive longitudinal data using ecological momentary assessment (EMA) methods, which aim to capture psychological, emotional, and environmental factors that may relate to a behavioral outcome in near real-time. In practice, JITAIs' effectiveness depends on accurately identifying high-risk situations by the user or by pre-determined decision rules to initiate the delivery of intervention components. Decision rules for efficacious interventions rely on a thorough understanding of the factors that characterize a subject's risk for a behavioral outcome, the dynamics of these risk factors' relation with the outcome over time, and the knowledge of possible strategies to target a risk factor \citep{nahum2017just}.
 
In the analysis of this paper, we investigate a behavioral health intervention study that targets smoking cessation.  Historically, smoking cessation studies have used health behavior theory \citep{shiffman2002immediate,timms2013dynamical} or group-level trends of smoking antecedents \citep{piasecki2013smoking} to determine when a JITAI should be triggered. However, this approach is limited since current health behavior models are inadequate for guiding the dynamic and granular nature of JITAIs \citep{riley2011health,klasnja2015microrandomized}. 
%\sout{For example, this practice may overlook important smoking lapse patterns in the data that may be useful for generating new hypotheses, as many health behavior theories were developed prior to the proliferation of real-time data collection methods. Also, static decision rules may fail to address within-person changes in behavior, since triggers for smoking lapse are likely to vary across individuals and within the same individual over time}  \citep{witkiewitz2011therapist,hendershot2011relapse}. 
Additionally, the design of efficacious smoking cessation interventions is challenged by the complexity of smoking behaviors around a quit attempt and misunderstandings of the addiction process \citep{piasecki2002have}. 
%For example, the relations between risk factors and smoking behaviors are often inconsistent across studies, potentially due to differences in the duration of the assessment period and assessment times} \citep{wray2013systematic, piasecki2006relapse}.  
More recently, smoking behavior researchers have capitalized on the ability of mHealth techniques to collect rich streams of data capturing subjects' experiences close to their occurrence at a high temporal resolution. The structure, as well as the complexity, of these data provide unique opportunities for the development and implementation of more advanced analytical methods compared to traditional longitudinal data analysis methods used in behavioral research (e.g., mixed models, growth curve models) \citep{trail2014functional}. For example, researchers have applied reinforcement learning \citep{luckett2019estimating} and dynamic systems approaches \citep{trail2014functional,rivera2007using, timms2013dynamical} to design and assess optimal treatment strategies using mHealth data. Additionally, \cite{koslovsky2018bayesian}, \cite{de2017use} and \cite{berardi2018markov} have applied hidden and observed Markov models to study transitions between discrete behavioral states, \cite{shiyko2012using} and \cite{dziak2015modeling} have used mixture models to identify latent structures, and \cite{kurum2016time} have employed joint modeling techniques to study the complexity of smoking behaviors. 
 
Greater insights into the dynamic relation between risk factors and smoking behaviors have been generated by the application of functional data techniques \citep{ trail2014functional,vasilenko2014time,koslovsky2017time, tan2012time}. These methods are well-suited for high-dimensional data with unbalanced and unequally-spaced observation times, matching the format of data collected with EMAs. They also require little assumptions on the structure of the relations between risk factors and behavioral outcomes. One popular approach uses varying-coefficient models, which belong to the class of generalized additive (mixed) models. These semiparametric regression models allow a covariate's corresponding coefficient to vary as a smooth function of other covariates \citep{hastie1993varying}. For example, \cite{selya2015nicotine} examined how the relation between the number of cigarettes smoked during a smoking event and smoking-related mood changes varies as a function of nicotine dependence. More frequently, penalized splines have been employed in varying-coefficient models to investigate how the effect of a covariate varies as a function of time, leading to time-varying effect models (TVEM) \citep{tan2012time,lanza2013advancing,koslovsky2017time, shiyko2012using, mason2015time,vasilenko2014time}. These approaches allow researchers to identify the critical moments that a particular risk factor is strongly associated with smoking behaviors, information that can be used to design tailored intervention strategies based on a subject's current risk profile. 

\subsection{Model Overview}

While there are various inferential challenges that functional data analysis models can address, in the application of this paper we focus on incorporating three recurring themes in behavioral research to explore the relations between risk factors and smoking behaviors:

\begin{enumerate}
	\item \textit{Model Assumptions} - Numerous smoking behavior research studies have relied on semiparametric, spline-based methods to learn the relational structure between risk factors and outcomes  \citep{tan2012time,vasilenko2014time}. 
	\item \textit{Variable Selection} - One of the main objectives of intensive longitudinal data analysis is to identify or re-affirm complex relations between risk factors and behavioral outcomes over time \citep{walls2005models}. 
	\item \textit{Latency} - A common aim in smoking behavior research studies is to identify latent structure in the data, such as groups or clusters of subjects with similar smoking behaviors over time \citep{mccarthy2016repeated, cursio2019latent, geiser2013analyzing, dziak2015modeling, brook2008developmental}. 
\end{enumerate}

To incorporate and expand upon these features in our analysis, we develop a flexible Bayesian varying-coefficient regression modeling framework for longitudinal binary responses that uses variable selection priors to provide insights into the dynamic relations between risk factors and outcomes. We embed spike-and-slab variable selection priors as mixtures of a point mass at zero (spike) and a diffuse distribution (slab) \citep{george1993variable,brown1998multivariate} and adopt the formulation of \cite{scheipl2012spike} to deconstruct the varying-coefficients terms, in our case time-varying effects, into a main effect, linear interaction term, and non-linear interaction term. Unlike previous approaches in behavioral health research that use time-varying effect models, our formulation allows us to gain inference on whether a given risk factor is related to the smoking behavior while also learning the type of relation. Additionally, by performing selection on fixed as well as random effects, our method is equipped to identify relations that vary over time and across subjects. For this, we exploit a P\'olya-Gamma augmentation scheme that enables efficient sampling without sacrificing interpretability of the regression coefficients as log odds ratios \citep{polson2013bayesian}. Furthermore, we adopt a Bayesian semiparametric approach to model fixed and random effects by replacing the traditional spike-and-slab prior with a nonparametric construction to cluster risk factors that have similar strengths of association.

\subsection{ Just-in-Time Adaptive Interventions for Smoking Abstinence}
Although multiple studies have examined momentary predictors of smoking lapse \citep{shiffman2000dynamic,piasecki2003smoking,businelle2014predicting}, JITAIs for smoking cessation are still nascent. Thus far, studies have used participant-labeled GPS coordinates to trigger supportive messages to prevent smoking \citep{naughton2016context}, or have tailored messages to the duration and intensity of participant's self-reported side effects while taking varenicline \citep{mcclure2016evaluating}. Using our proposed approach, we analyze ILD collected in a study investigating the utility of a novel, smartphone-based smoking cessation JITAI (\textit{SmartT}). The \textit{SmartT} intervention \citep{businelle2016ecological} uses a lapse risk estimator to identify moments of heightened risk for lapse, and tailors treatment messages in real-time based upon the level of imminent smoking lapse risk and currently present lapse triggers. To our knowledge, no other studies have used EMA data to estimate risk for imminent smoking lapse and deliver situation-specific, individually-tailored treatment content prior to lapse.
 
%Details of the study design are described elsewhere (Citation Redacted). 
In this study, adult smokers (N=81) recruited from a smoking cessation research clinic were randomized to the \textit{SmartT} intervention, the National Cancer Institute's QuitGuide (\textit{NCI QuitGuide}), or weekly counseling sessions (\textit{usual care}), and followed over a five-week period spanning one week prior to a scheduled quit attempt to four weeks after. At the beginning of the assessment period, baseline measures were collected, and subjects were shown how to complete EMAs on a study-provided smartphone. Throughout the assessment period, subjects completed daily diaries and received four random EMAs from the smartphone to complete each day. For each EMA, subjects were prompted on their recent smoking behaviors, alcohol consumption, as well as various questions regarding their current psychological, social, and environmental factors that may contribute to an increased risk of smoking behaviors. 

%We analyze the relations between momentary smoking (defined as whether or not a subject reported smoking in the last 4 hours) and potential baseline and EMA items. 
Findings indicate that our approach is well-positioned to help researchers evaluate, design, and deliver tailored intervention strategies in the critical moments surrounding a quit attempt. In particular, results confirm previously identified temporal relations between smoking behaviors around a quit attempt and risk factors.
% without making any model assumptions regarding their presence or structure. 
They also indicate that subjects differ in how they respond to different risk factors over time. Furthermore, we identify clusters of active risk factors that can help researchers prioritize intervention strategies  based on their relative strength of association at a given moment.  Importantly, our approach generates these insights with minimal assumptions regarding which risk factors were related to smoking in the presence of others, the structural form of the relation for active terms, or the parametric form of regression coefficients. %Additionally, our method borrows strength across regression coefficients to efficiently provide estimates for individual components of the smooth functions and random effects, each providing different pieces of information that can be used to design tailored intervention strategies.
 
The rest of the paper is organized as follows. In section \ref{sec:methods}, we present our modeling approach and describe prior constructions. In section \ref{sec:case}, we investigate the relations between risk factors and smoking behaviors in the critical moments surrounding a scheduled quit attempt using mHealth data. In section \ref{sec:simul}, we conduct a simulation study investigating the variable selection and clustering performance of our proposed method on simulated data. In section \ref{sec:sens}, we evaluate prior sensitivity of our model.  In section \ref{sec:remarks}, we provide concluding remarks.

\section{Methods}
\label{sec:methods}
The objective of our analysis is to identify relations between a set of risk factors (i.e., baseline and EMA items) and a binary outcome (i.e., momentary smoking) repeatedly collected over time. For this, we employ a Bayesian variable selection framework that allows a flexible structure for the unknown relations. We achieve this by performing selection not only on main effects, but additionally on linear and non-linear interaction terms as well as random effects. In this work, we refer to fixed and random effects in the context of hierarchical or multilevel models, where fixed effects are constant across subjects and random effects differ at the subject-level. We chose this terminology based on its familiarity within both frequentist and Bayesian paradigms, but point out that the fixed or population-level effects are treated as random variables in our model, and thus follow a probability distribution. 
%Instead of naively assuming that these components are unrelated, we further impose nonparametric priors which reduce the parameter space by implicitly clustering risk factors' corresponding regression coefficients. 

\subsection{A Varying-Coefficient Model for Intensive Longitudinal Data Collected with EMAs}
Let $y_{ij} \in \{0,1\}$ represent momentary smoking   for subject $i = 1,\dots, N$, and $ \boldsymbol{x}_{ij}$ and $\boldsymbol{z}_{ij} $ represent $P$- and $D$-dimensional vectors of risk factors collected on each subject at time $j = 1,\dots, n_i$, respectively. To maintain temporality in our particular application (see section \ref{sec:case} for more details), we model the relation between momentary smoking by the next assessment and current, potential risk factors as a varying-coefficient model of the type 

\begin{equation}\label{VICM}
logit(P(y_{i,j+1} = 1|\boldsymbol{x}_{ij},\boldsymbol{z}_{ij},u_{ij})) = \sum_{p=1}^{P}f_p(u_{ij}) x_{ijp} + \boldsymbol{\alpha}^{\prime}_i\boldsymbol{z}_{ij},
\end{equation}  
where $f_p(u)$ are smooth functions of a scalar covariate $u$, and $\boldsymbol{\alpha}_i$ represents subject specific random effects. Similar temporal assumptions have been made previously in smoking behavior research studies \citep{bolman2018predicting,minami2014relations, shiffman1996first,shiffman2013conceptualizing,shiyko2014modeling}. Note that in general, researchers may use the framework of \ref{VICM} to model the relation between a binary outcome and potential risk factors collected concurrently, in addition to lagged trends, as is typical in longitudinal studies  \citep{fitzmaurice2012applied}. With this formulation, we include varying-coefficient terms for each of the $P$ risk factors based on $u$. However in general, we can specify varying-coefficient terms that depend on $u^{\prime} \ne u$, and thus the number of varying-coefficient terms in the full model is not strictly $P$. If $u$ is chosen to represent time, then this model is commonly referred to as a \textit{time-varying effect model} in smoking behavior research \citep{tan2012time,vasilenko2014time,dziak2015modeling,koslovsky2017time}. Note that $ \boldsymbol{z}_{ij}$ is typically a subset of $\boldsymbol{x}_{ij} $ \citep{kinney2007fixed,cheng2010real,hui2017hierarchical} and that incorporating a 1 in $ \boldsymbol{x}_{ij}$ and $\boldsymbol{z}_{ij} $, allows for an intercept term that varies as a function of $u$ and a random intercept term, respectively.  Additionally, this formulation can handle time-invariant risk factors, such as baseline items, by fixing $x_{ijp}$ ($z_{ijd}$) to $x_{ip}$ ($z_{id}$) for all observations $j$.

We approximate the smooth functions with spline basis functions. Specifically,
\begin{equation}\label{smooth}
 f_p(u_{ij})= \boldsymbol{ \mathcal{U}}_{ij}^{\prime} \boldsymbol{\phi}_p,
\end{equation}
where $\boldsymbol{ \mathcal{U}}_{ij}$ is a spline basis function for $u_{ij}$, and $ \boldsymbol{\phi}_p$ is a $r_p$-dimensional vector of corresponding spline coefficients. For simplicity, the splines are constructed with an equal number of equally spaced knots that depend on the minimum and maximum values of $\boldsymbol{u}$.

\subsection{Penalized Priors for the Spline Coefficients}\label{Spline}

Using a combination of variable selection and shrinkage priors, our approach generates insights on the underlying structure of the smooth functions by reconstructing them as the summation of main effect, linear interaction, and non-linear interaction components. Formally, we rewrite Equation (Eq.) \eqref{smooth} as
\begin{equation}\label{smoothII}
 f_p(u_{ij}) = \beta^{*}_p\boldsymbol{ \mathcal{U}}_{ij}^{*\prime}\boldsymbol{\xi}_p + \beta_p^{\circ}u_{ij} + \beta_{0p},	
\end{equation}
where the constant term $\beta_{0p}$ captures the main effect of $\boldsymbol{x}_p$, $\beta_p^{\circ}$ represents the effect of the linear interaction between $\boldsymbol{u}$ and $\boldsymbol{x}_p$, and $\beta^{*}_p\boldsymbol{\xi}_p$ is a parameter-expanded vector of coefficients corresponding to the non-linear interaction term. 

To derive the non-linear component in Eq. \eqref{smoothII}, we start by penalizing the spline functions in Eq. \eqref{smooth} with a second-order Gaussian random walk prior following 
\begin{equation}\label{smoothprior}
\boldsymbol{ \mathcal{U}} \boldsymbol{\phi}_p |s^2 \sim N(\boldsymbol{0},s^2\boldsymbol{ \mathcal{U}}\boldsymbol{P}^{-}\boldsymbol{ \mathcal{U}}^{\prime}),
\end{equation} 
where  $\boldsymbol{\mathcal{U}}$ is a $\sum_{i = 1}^N (n_i-1) \times r_p$-dimensional matrix with each row corresponding to $\boldsymbol{\mathcal{U}}_{ij}^{\prime}$ for the $i^{th}$ subject at the $j^{th}$ assessment, $s^2$ controls the amount of smoothness, and $\boldsymbol{P}$ is the appropriate penalty matrix \citep{lang2004bayesian}. Next, we take the spectral decomposition of  $\boldsymbol{ \mathcal{U}}\boldsymbol{P}^{-}\boldsymbol{ \mathcal{U}}^{\prime} =$
$\begin{bmatrix}
\boldsymbol{U}_+ & \boldsymbol{U}_{\circ} \end{bmatrix}$ 
$\begin{bmatrix}
\boldsymbol{V}_+ & \boldsymbol{0}\\
\boldsymbol{0} & \boldsymbol{0}
\end{bmatrix}$ 
$\begin{bmatrix}
\boldsymbol{U}_+  \\
\boldsymbol{U}_{\circ}\end{bmatrix},$ where $\boldsymbol{U}_+$ is a matrix of eigenvectors with corresponding positive eigenvalues along the diagonal of matrix $\boldsymbol{V}_+$, and $\boldsymbol{U}_{\circ}$ are the eigenvectors associated with the zero eigenvalues. Now, we can re-define the smooth functions in Eq. \eqref{smooth} as the sum of non-linear (penalized) interaction, linear (non-penalized) interaction, and main effect terms as presented in Eq. \eqref{smoothII}, where the penalized term is written as $\boldsymbol{\mathcal{U}}^*\boldsymbol{\varphi}_p^*$ with $\boldsymbol{\mathcal{U}}^* = \boldsymbol{U}_+\boldsymbol{V}_+^{1/2}$. By assuming independent normal priors for  $\boldsymbol{\varphi}_p^*$, a proper prior for the penalized terms that is proportional to Eq. \eqref{smoothprior} can be obtained. 

We take two additional measures to enhance the computational efficiency of the resulting MCMC algorithm. First, only eigenvalues/vectors that explain a majority of the variability in Eq. \eqref{smoothprior} are used to construct $\boldsymbol{\mathcal{U}}^*$. Additionally, we apply a parameter-expansion technique for the penalized terms in $f_p
 (\cdot)$, setting $\boldsymbol{\varphi}_p^* = \beta^{*}_p\boldsymbol{\xi}_p$, where $\beta^{*}_p$ is a scalar and $\boldsymbol{\xi}_p$ is a vector with the same dimension as $\boldsymbol{\varphi}_p^*$. This technique enables us to perform selection on the penalized terms as a group rather than determining their inclusion separately. By rescaling $\beta^{*}_p$ and $\boldsymbol{\xi}_p$ at each MCMC iteration, such that $|\boldsymbol{\xi}_p|$ has mean equal to one, $\boldsymbol{\xi}_p$ maintains the shape of the smooth function and $\beta^{*}_p$ represents the term's strength of association, while preserving identifiability, similar to  \cite{scheipl2012spike}.

For variable selection, we impose spike-and-slab prior distributions on the $3*P = T$-dimensional vector $\boldsymbol{\beta} = (\beta^*_1,\beta^{\circ}_1,\beta_{01},\dots, \beta^*_P,\beta^{\circ}_P, \beta_{0P})^{\prime}$. In general, the spike-and-slab prior distribution is composed of a mixture of a Dirac delta function at zero, $\delta_0(\cdot)$, and a known distribution, $\mathcal{S}(\cdot)$, such as a normal with mean zero and diffuse variance \citep{george1993variable,brown1998multivariate}. A latent indicator variable, $\nu_t$, representing a risk factor's inclusion or exclusion in the model determines whether the risk factor's regression coefficient is set to zero (spike) or free to be estimated in the model (slab). Specifically for a given coefficient $\beta_t$, we assume 
\begin{eqnarray}
\label{eq:prior_beta}
\beta_t|\nu_t \sim \nu_t\cdot \mathcal{S}(\beta_t) + (1-\nu_t)\delta_0(\beta_t).
\end{eqnarray}

To complete the prior specification for this portion of the model, we assume that the slab component, $\mathcal{S}(\beta_t)$, follows a $N(0,\tau^2)$ with variance $\tau^2$, and that the inclusion indicators are distributed as $\nu_t|\theta_t \sim \mbox{Bernoulli}(\theta_t)$, with prior probability of inclusion $\theta_t \sim \mbox{Beta}(a_{\nu_t},b_{\nu_t})$. Integrating out $\theta_t$ we obtain $\nu_t \sim$ Beta-Binomial$(a_{\nu_t},b_{\nu_t})$, where hyperparameters $a_{\nu_t}$ and $b_{\nu_t}$ are set to control the sparsity in the model. Lastly, each element of $\boldsymbol{\xi}_p $, $\xi_{pr}$, is assumed to follow a $N(\mu_{pr},1)$, with mean $\mu_{pr} =\pm1$ with equal probability.  
Placing a majority of the prior mass for each $ \xi_{pr}$ around $\pm 1$ is motivated by the role it plays in the expansion of $\boldsymbol{\varphi}_p^*$, as described above. 

\subsection{Prior Specification for the Random Effects}

We perform selection on the random effects, $\boldsymbol{\alpha}_i$, using the modified Cholesky decomposition approach of \cite{chen2003random}. Specifically, we reparameterize the random effects
\begin{equation}
\boldsymbol{\alpha}_{i} = \boldsymbol{K}\boldsymbol{\Gamma}\boldsymbol{\zeta}_i,
\end{equation}
where $\boldsymbol{K}$ a positive diagonal matrix with elements $\boldsymbol{\kappa} = (\kappa_1,\dots, \kappa_D)^{\prime}$, and $\boldsymbol{\Gamma}$ a lower triangle matrix with diagonal elements set to one and free elements otherwise. To perform variable selection, we set the prior for $\boldsymbol{\kappa}$ to follow a similar spike-and-slab prior distribution as in section 2.2, where the slab distribution $\mathcal{S}(\kappa_d) = FN(m_0,v_0)$. Here, $FN$ represents a folded normal distribution defined as $$FN(m_0,v_0) =  (2\pi v_0)^{-1/2}\exp(-(\kappa_d - m_0)^2/(2v_0)) + (2\pi v_0)^{-1/2}\exp(-(\kappa_d + m_0)^2/(2v_0)),$$ where  $m_0 \in \mathbb{R}$ and $v_0 > 0 $ are location and scale parameters, respectively. Note that we forgo the parameter-expansion approach of \cite{kinney2007fixed}, which introduces a redundant multiplicative parameter in the implied random effect covariance matrix, in favor of a model that enables meaningful inference for $\boldsymbol{\kappa}$ and ultimiately their cluster assignments. Similar to section \ref{Spline}, we let the corresponding inclusion indicators $\lambda_d$ follow a Beta-Binomial$(a_{\lambda_d},b_{\lambda_d})$ to induce sparsity on the random effect terms. Lastly, we assume the $D(D-1)/2$-dimensional vector of free elements in $\boldsymbol{\Gamma}$ follow $N(\boldsymbol{\gamma}_0, V_{\gamma}) \cdot I(\boldsymbol{\gamma} \in \mathcal{Z}) $, where $I$ represents an indicator function, and $\mathcal{Z}$ represents the parameters with corresponding random effects included in the model. For example, if the $d^{th}$ random effect is included (i.e., $\lambda_d = 1$), then  $\gamma_{d1},\dots,\gamma_{d,d-1}\mbox{ and } \gamma_{d+1,d}, \dots \gamma_{D,d} \in \mathcal{Z}$. Lastly, we assume  $\boldsymbol{\zeta}_i \sim N(\boldsymbol{0},\boldsymbol{I}).$

\subsection{Spiked Nonparametric Priors} To complete our approach, we  investigate nonparametric prior constructions for the spike-and-slab components of the reparameterized fixed and random effects by assuming that the slab component follows a Dirichlet process (DP). These priors are commonly referred to as spiked DP (SDP) priors \citep{canale2017pitman,kim2009spiked,savitsky2010spiked,dunson2008bayesian}. In the context of our model, SDP priors allow us to simultaneously select influential risk factors while clustering effects with similar relations to the smoking outcome. The formulation we use here is sometimes refers to as an ``outer" SDP prior, since the point mass at zero is outside of the base distribution of the DP. Alternatively, the ``inner" construction places the spike-and-slab prior inside the DP, serving as the base distribution.  The inner formulation provides the opportunity for coefficients to cluster at zero, but does not force a point mass at zero explicitly. As such, the likelihood that a coefficient is assigned to the trivial cluster grows with the number of coefficients excluded from the 
 model. Alternatively,  the outer formulation is a more informative prior, since it explicitly assigns a point mass at zero, and, in addition, carries less computational demands since it does not require auxiliary variables for MCMC sampling \citep{neal2000markov, savitsky2010spiked}. We refer readers to \cite{canale2017pitman} for a detailed explanation of the structural differences between the two prior formulations. 

%A Dirichlet process, $DP(\alpha_0,G_0)$, is defined as the distribution of a random probability measure $G$ over the measurable space $(\Omega,\mathcal{B})$ such that for any finite partition $(B_1,\dots B_m)$ of $\Omega$, the random vector $(G(B_1),\dots ,G(B_m)) \sim \mbox{Dirichlet}(\alpha_0G_0(B_1), \dots, \alpha_0G_0(B_m))$ \citep{ferguson1973bayesian}. $G_0$ and $\alpha_0$ are referred to as the base distribution and concentration parameter, respectively.  Following the P\'olya urn characterization \citep{blackwell1973ferguson}, it is easily seen that the DP is discrete with probability one, implying a positive probability that coefficients share similar effect sizes. For illustration, assume that $(\beta_1, \dots, \beta_{P-1})$ is an iid sample from $G$, the predictive distribution for the next coefficient is described as $\beta_P | \beta_1 ,\dots, \beta_{P-1} \sim \sum_{l = 1}^{P-1}\frac{1}{\alpha_0 + P - 1}\delta_{ \beta_l} + \frac{\alpha_0}{\alpha_0 + P - 1 }G_0,$ where $\delta_{ \beta_l}$ is a Dirac delta measure at $ \beta_l$. 

First, we assume the regression coefficients associated with the main effects and linear interaction terms follow a SDP to provide insights on risk factors that share underlying linear trends with momentary smoking by the next assessment over the course of the study. Specifically, we assume the slab component in Eq. \eqref{eq:prior_beta}  is a Dirichlet process prior $H \sim DP(\vartheta,H_0)$, with base distribution $H_0 = N(0,\tau^2)$ and
concentration parameter $\vartheta$. Furthermore, we assume a hyperprior $\vartheta \sim G(a_{\vartheta},b_{\vartheta})$, with $a_{\vartheta},b_{\vartheta} > 0$. For the nonlinear interaction terms, we avoid the SDP since it would produce uninterpretable cluster assignments due to the parameter-expansion approach taken to improve selection performance. For example, similar values for $\beta_t^*$ and $\beta_{t'}^*$ may correspond to vastly different $\boldsymbol{\varphi}^*_t$ and $\boldsymbol{\varphi}^*_{t'}$, depending on their respective $\boldsymbol{\xi}$ and spline basis functions. Similarly, placing a DP prior on the individual components in $\boldsymbol{\xi}$, or even $\boldsymbol{\varphi}$, would not provide interpretable results on the overall nonlinear effect. 
We take a similar approach for the random effects. Here, we assume the slab components for the diagonal elements of $\boldsymbol{K}$, $\mathcal{S}(\kappa_d) = W$, $W \sim DP( \mathcal{A}, W_0),$ where $W_0 \sim FN(m_0,v_0)$, and $\mathcal{A}$ is the concentration parameter of the DP. To complete the prior assumptions for the random effects portion of the model, let $\mathcal{A} \sim G(a_{\mathcal{A}}, b_{\mathcal{A}})$, where $a_{\mathcal{A}},b_{\mathcal{A}}>0$ are shape and rate parameters, respectively. 

There is evidence that relaxing parametric assumptions for random effects using DP priors may cause inferential challenges as the mean of the random effects are non-zero almost surely \citep{li2011center,yang2012bayesian,cai2017bayesian}. Our approach differs in that we do not directly replace the typical normal assumption for random effects with a nonparametric prior. Instead, we place a nonparametric prior on the covariance decomposition components, $\boldsymbol{K}$, while letting $\boldsymbol{\zeta}_i$ follow a normal distribution centered at zero.  
%\textit{Proposition 1} - 
%If  $\boldsymbol{\zeta}_i$ is independent of $\boldsymbol{K\Gamma}$, then $E[\boldsymbol{K\Gamma\zeta}_i] = \boldsymbol{0}$. \\
As such, our approach avoids any identifiability issues with the fixed effects while still relaxing the  parametric assumption on the reparameterized random effects, $\boldsymbol{K \Gamma \zeta}_i$. It is important to note that by doing this we are adopting a Bayesian semiparametric modeling structure, since the random effects are linear combinations of spiked Dirichlet process and normal random variables  \citep{muller2007semiparametric}.

%A graphical representation of the extended version of our model formulation is displayed in Figure \ref{fig:graph}.

%\begin{figure}
%\centering
%\includegraphics[scale=0.8]{DAG_Applied}
%\caption{Graphical representation of the proposed model with spiked nonparametric priors. Note that hyperparameters are suppressed for clarity. }
%\label{fig:graph}
%\end{figure}

\subsection{Posterior Inference}
For posterior inference, we implement a Metropolis-Hastings within Gibbs algorithm. The full joint model is defined as 
$$  f(\boldsymbol{y}|\boldsymbol{\varrho},\boldsymbol{\omega},\boldsymbol{x},\boldsymbol{u},\boldsymbol{z})p(\boldsymbol{\omega})p(\boldsymbol{\beta}|\boldsymbol{\nu})p(\boldsymbol{\nu})p(\vartheta)p(\boldsymbol{K}|\boldsymbol{\lambda})p(\boldsymbol{\lambda})p(\mathcal{A})p(\boldsymbol{\xi}|\boldsymbol{\mu})p( \boldsymbol{\mu})p(\boldsymbol{\zeta})p(\boldsymbol{\Gamma}),$$
where $\boldsymbol{\varrho} = \{\boldsymbol{\beta},  
\boldsymbol{\xi}, \boldsymbol{K}, \boldsymbol{\Gamma},\boldsymbol{\zeta} \}$. 
We use the P\'olya-Gamma augmentation of \cite{polson2013bayesian} to efficiently sample the posterior distribution for the logistic regression model. Following  \cite{polson2013bayesian}, we express the likelihood contribution of $y_{i,j+1}$ as  
$$f(y_{i,j+1}|\cdot) = \frac{(e^{\psi_{ij}})^{y_{i,j+1}}}{(1 + e^{\psi_{ij}})} \propto \exp({k_{i,j+1}\psi_{ij}})\int_{0}^{\infty}\exp(-\omega_{i,j+1}\psi_{ij}^2/2)p(\omega_{i,j+1}|n_{i,j+1},0)\partial \omega, $$
where $k_{i,j+1} = y_{i,j+1} - n_{i,j+1}/2$, $p(\omega_{i,j+1}|n_{i,j+1},0) \sim PG(n_{i,j+1},0)$, and $PG$ is the P\'olya-Gamma distribution. Using the notation presented in the previous sections, we set
$$\psi_{ij} =  \sum_{p=1}^{P}(\beta^{*}_p\boldsymbol{ \mathcal{U}}_{ij}^*\boldsymbol{\xi}_p + \beta_p^{\circ}u_{ij} + \beta_{0p})x_{ijp} + \boldsymbol{z}_{ij}^{\prime} \boldsymbol{K} \boldsymbol{\Gamma} \boldsymbol{\zeta}_i.$$

The MCMC sampler used to implement our model is outlined below in Algorithm 1. A more detailed description of the MCMC steps as well as a graphical representation of the model are provided in the Supplementary Material. After burn-in and thinning, the remaining samples obtained from running Algorithm 1 for $\tilde{T}$ iterations are used for inference. To determine a risk factor's inclusion in the model, its marginal posterior probability of inclusion (MPPI) is empirically estimated by calculating the average of its respective inclusion indicator's MCMC samples \citep{george1997approaches}.  Note that inclusion for both fixed and random effects is determined marginally for $\beta_t$ and $\lambda_d$, respectively. Commonly, covariates are included in the model if their MPPI exceeds 0.50 \citep{barbieri2004optimal} or a Bayesian false discovery rate threshold, which controls for multiplicity \citep{newton2004detecting}.

\begin{algorithm} 
	\caption{MCMC Sampler}\label{MCMC}
%	\small
	\begin{algorithmic}[1]
		\State Input data $\boldsymbol{y},\boldsymbol{x},\boldsymbol{u},\boldsymbol{z}$
		\State Initialize parameters: $\boldsymbol{\varrho}, \boldsymbol{\omega}, \boldsymbol{\nu},\boldsymbol{\lambda},\vartheta, \mathcal{A}, \boldsymbol{\mu}$
		\State Set $DP_{\bar{\boldsymbol{\beta}}}$ and $DP_{\boldsymbol{K}}$ to True or False to indicate DP for slab on fixed or random effects, respectively.
		\For{iteration $\tilde t = 1,\dots,\tilde T$}
		\For{$i = 1,\dots,N$}
		\For{ $j = 1,\dots,n_i-1$}
		\State Update $\omega_{i,j+1} \sim PG(1,\psi_{ij})$
		\EndFor
		\EndFor
		\If {$DP_{\bar{\boldsymbol{\beta}}}$ }
		\State Update cluster assignment of $\bar{\boldsymbol{\beta}}$ following  \cite{neal2000markov} algorithm 2.
		\EndIf
		\State Jointly update $\boldsymbol{\beta}$ and $\boldsymbol{\nu}$ with Between and Within Step following \cite{savitsky2011variable}.
		\State Update $\boldsymbol{\xi}$ from FCD $N(\mu_{\boldsymbol{\xi}},V_{\boldsymbol{\xi}})$.
		\For{ $p = 1,\dots,P$}
		\State Rescale $\boldsymbol{\xi}_p^*$ and $\beta_p^*$ so $\boldsymbol{\varphi}_p^*$ remains unchanged.
		\EndFor
		\For{$p = 1,\dots,P$}
		\For{ $r = 1,\dots,r_p$}
		\State Set $\mu_{pr} = 1$ with probabilty  $1/(1 + \exp(-2\xi_{pr}))$.
		\EndFor
		\EndFor
		\State Update $\vartheta$ by the two-step Gibbs update of \cite{escobar1995bayesian}.
		\If {$DP_{\boldsymbol{K}}$ }
		\State Update cluster assignment of $DP_{\boldsymbol{K}}$ following  \cite{neal2000markov} algorithm 2. 
		\EndIf
		\State Jointly update $\boldsymbol{K}$ and $\boldsymbol{\lambda}$ with Between and Within Step following \cite{savitsky2011variable}.
		\State Update $\mathcal{A}$ following two-step Gibbs update of \cite{escobar1995bayesian}.
		\State Update $\boldsymbol{\Gamma}$ from FCD $N(\hat{\boldsymbol{\gamma}},\hat{V}_{\gamma})\cdot I(\boldsymbol{\gamma} \in \mathcal{Z})$.
		\For{ $i = 1,\dots, N $ }
		\State Update $\boldsymbol{\zeta}_i$ from FCD $N(\hat{\boldsymbol{\zeta}}_i,\hat{V}_{\zeta_i})$. \EndFor
		\EndFor
	\end{algorithmic}
\end{algorithm} 

\section{Case Study}
\label{sec:case}

In this section, we study the smoking behaviors in a group of adult smokers recruited from a smoking cessation research clinic. The overall research goal of this study was to identify and investigate the structural form of the relations between a set of risk factors and smoking over a five-week period surrounding a scheduled quit attempt, using intensive longitudinal data collected with EMAs. 

\subsection{Data Analysis}
In the study design, momentary smoking, our outcome of interest, was defined as whether or not a subject reported smoking in the 4 hours prior to the current EMA. However at each EMA, a subject was prompted on their \textit{current} psychological, social, environmental, and behavioral status. Thus to maintain temporality in this study, we assessed the relations between momentary smoking and measurements collected in the previous EMA. As such, regression coefficients are interpreted as the log odds of momentary smoking by the next assessment for a particular risk factor. In this study, we investigated psychological and affective factors including \textit{urge} to smoke, feelings of \textit{restlessness}, \textit{negative affect} (i.e., irritability, frustration/anger, sadness, worry, misery), \textit{positive affect} (i.e., happiness and calmness), being \textit{bored}, \textit{anxiousness}, and \textit{motivation to quit smoking}. Additionally, we investigated numerous social and environmental factors such as whether or not the subject was \textit{interacting with a smoker}, if cigarettes were easily available (\textit{cigarette availability}), and whether or not the subject was drinking alcohol (\textit{alcohol consumption}). Also, we included a set of baseline, time-invariant measures (i.e., heaviness of smoking index (\textit{HSI}), \textit{age} (years), being \textit{female}, and treatment assignment) into the model. For each of these risk factors, we included a fixed main effect, linear interaction, and non-linear interaction term as well as a random main effect and linear interaction term. All interactions investigated in this analysis were between risk factors and assessment time (i.e., $u_{ij} = t_{ij}$), and $t_{ij}$ were centered so that $t=0$ represents the beginning of the scheduled quit attempt.

 Only complete EMAs with corresponding timestamps were included in this analysis, resulting in 9,634 total observations with the median number of assessments per individual 151 (IQR 101.5-162). All continuous covariates were standardized to mean zero and variance one before analysis to help reduce multicollinearity and place covariates on the same scale for interpretation. The spline functions were initially generated with 20 basis functions, but only the eigenvalues/eigenvectors that captured 99.9\% of the variability were included in the model to reduce the parameter space and computation time, similar to \citep{scheipl2012spike}. This reduced the column space of the penalized covariates $\boldsymbol{\mathcal{U}}^*$ to 8 in our application. We applied our model with the traditional spike-and-slab prior, as well as the spiked DP. When fitting each model,
we chose a non-informative prior for the fixed and random effects' inclusion indicators,  $a_{\nu_t} = b_{\nu_t} =  a_{\lambda_d} =  b_{\lambda_d} = 1$. This assumption reflects the exploratory nature of our study aimed at learning potential relations between risk factors and smoking behaviors with little or no information regarding their occurrence in the presence of other risk factors. We assumed a mildly informative prior on the fixed regression coefficients by setting $\tau^2= 2$. This places a 95\% prior probability of included regression coefficients between an odds ratio of 0.06 and 16. Additionally, we set
$ v_0= v^*= 10$, $m_0 = m^* = 0$, and ${\Gamma} \sim N( \boldsymbol{\gamma}_0 = \boldsymbol{0},\boldsymbol{V}_{\gamma}=\boldsymbol{I} )$. Lastly, when using the SDP prior, the hyperparameters for the concentration parameters $\vartheta$ and $\mathcal{A}$ were set to $a_{\vartheta} = b_{\vartheta} = a_{\mathcal{A}} = b_{\mathcal{A}} = 1$. For posterior inference, we ran our MCMC algorithm with and without SDP priors for both fixed and random effects for 10,000 iterations, treating the first 5,000 as burn-in and thinning to every 10$^{th}$ iteration. Trace plots of the parameters' posterior samples indicated good convergence and mixing. Additionally, we observed a relatively high correlation ($\sim 97$\%) between the posterior probabilities of inclusion obtained from two chains initiated with different parameter values, and potential scale reduction factors, $\hat{R}$, for each of the selected $\boldsymbol{\beta}$ and $\boldsymbol{K}$ below 1.1 \citep{gelman1992inference}, further demonstrating that the MCMC procedure was working properly and the chains converged. To assess model fit, a residual plot and a series of posterior predictive checks were performed in which we compared replicated data sets from the posterior predictive distribution of the model to the observed data \citep{gelman2000diagnostic}. Overall, we found strong evidence of good model fit. See the Supplementary Materials for details.  Inclusion in the model was determined using the median model approach \citep{barbieri2004optimal} (i.e., marginal posterior probability of inclusion (MPPI) $\geq 0.50 $). For the SDP model, clusters of regression coefficients were determined using sequentially-allocated latent structure optimization to minimize the lower bound of the variation of information loss \citep{wade2018bayesian,dahl2017}. To compare the predictive performance of both models, we performed a leave-one-out cross-validation approximation procedure, following the approach proposed by \cite{vehtari2017practical}. This approach approximates leave-one-out (LOO) cross-validation with the expected log pointwise predictive density (epld). By using Pareto smoothed importance sampling (PSIS) for estimation, it provides a more stable estimate compared to the method of \cite{gelfand1996model}. We used the R package \texttt{loo} \citep{vehtari2016loo},  which requires the pointwise log-likelihood for each subject $i = 1, \dots , N$ at each observation $j = 1, \dots, n_i$ calculated at each MCMC iteration $s = 1, \dots, S$, and produces an estimated $\widehat{\mbox{epld}}$ value, with larger values implying a superior model.

\subsection{Results}
\begin{figure}
	\centering
	\includegraphics[scale=0.39]{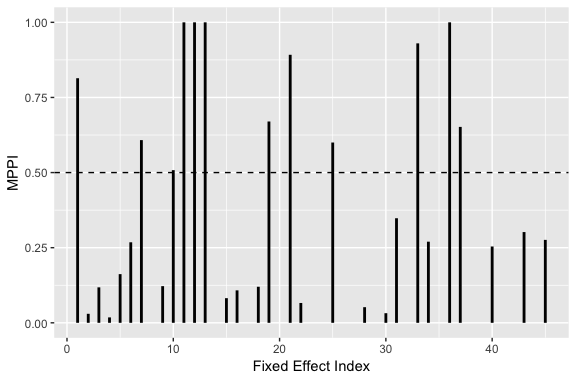}
	\includegraphics[scale=0.39]{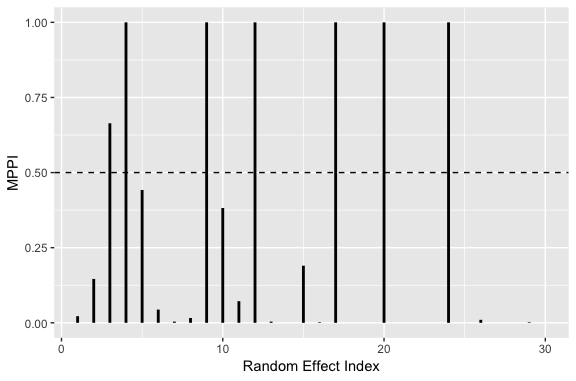}
	\caption{Smoking Cessation Study: Marginal posterior probabilities of inclusion (MPPI) for fixed ({\it top}) and random ({\it bottom}) effects. Selected fixed effects in ascending order: NCI (NL-INTX), urge to quit (NL-INTX), cigarette availability (all), interacting with a smoker (NL-INTX), negative affect (NL-INTX, main), being bored (NL-INTX), alcohol consumption (main), motivation to quit (main), HSI (NL-INTX). Selected random effects in ascending order: urge (main), cigarette availability (main), being bored (main), motivation to quit (main), SmartT (L-INTX), interacting with a smoker (L-INTX), being bored (L-INTX). Dotted lines represent the inclusion threshold of 0.50. NL-INTX: non-linear interaction, L-INTX: linear interaction }
	\label{fig:MPPI}
\end{figure}

\begin{figure}
	\centering
	\includegraphics[scale=0.3]{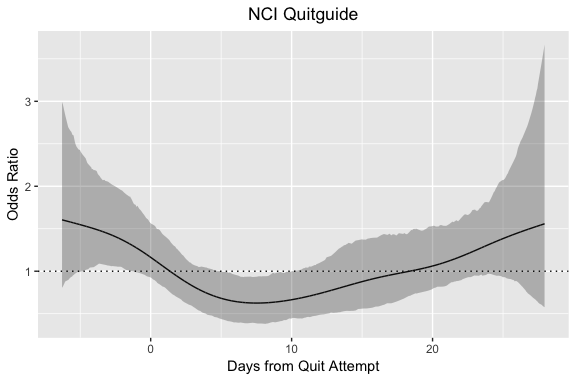}
	\includegraphics[scale=0.3]{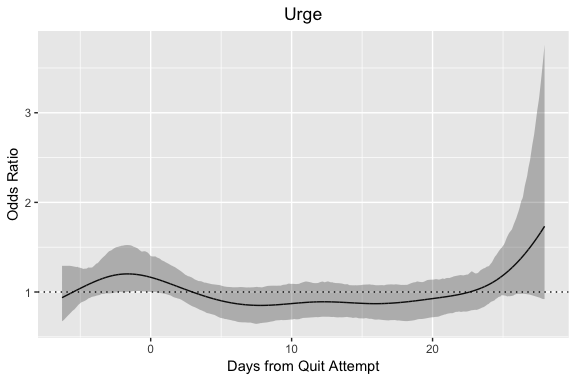}\\
	\vskip 2mm
	\includegraphics[scale=0.3]{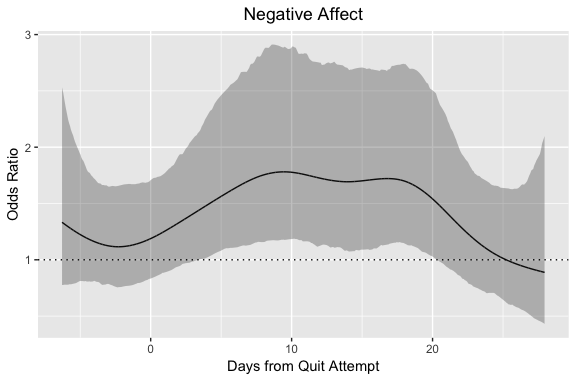}
	\includegraphics[scale=0.3]{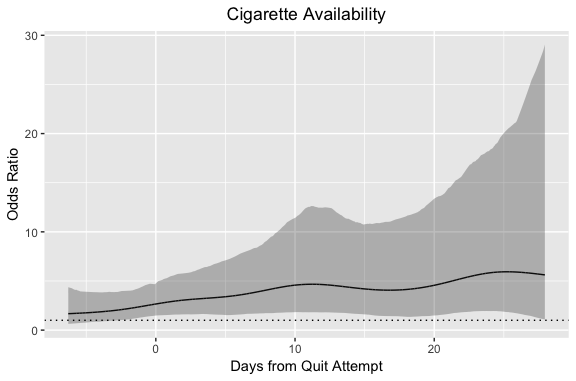}\\
	\vskip 2mm
	\includegraphics[scale=0.3]{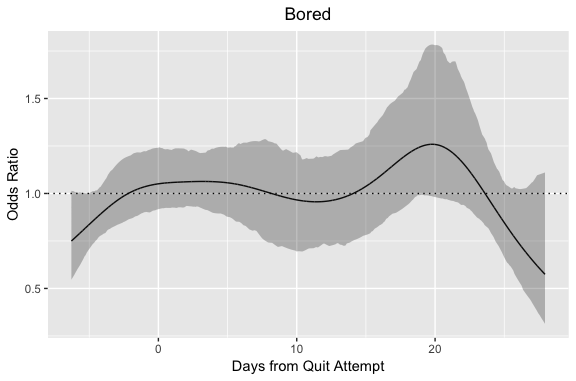}
	\includegraphics[scale=0.3]{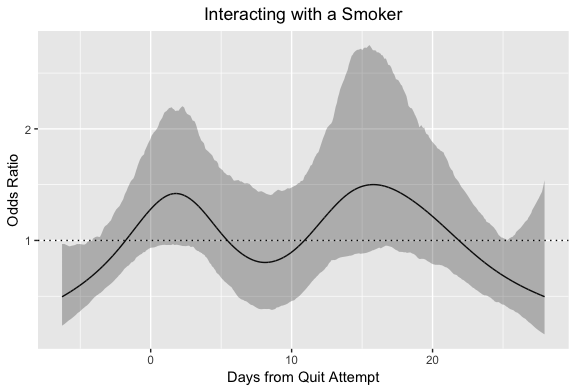}	
		\caption{Smoking Cessation Study: Time-varying effects on momentary smoking by the next assessment of those covariates selected by our model with SDP priors. Shaded regions represent pointwise 95\% CI. Dashed lines indicate an odds ratio of one. } 
	\label{fig:curves}
\end{figure}

 Overall, we found better predictive performance for the model with SDP priors versus the traditional spike-and-slab priors, 
$\widehat{\mbox{epld}}_{SDP} = -2985.1 $ and $\widehat{\mbox{epld}}_{SS} = -3062.7 $, respectively.  Plots of the marginal posterior probabilities of inclusion for the fixed and random effects selected using our proposed approach with SDP priors are found in Figure \ref{fig:MPPI}. Figure \ref{fig:curves} presents the time-varying effects selected using the same model. Compared to \textit{usual care}, we found a higher odds of momentary smoking by the next assessment for those assigned to the \textit{NCI QuitGuide} group prior to the quit attempt. However immediately after the quit attempt, we observed a lower odds of momentary smoking by the next assessment for those assigned to the \textit{NCI QuitGuide} group, which gradually increased to the initial level over the remainder of the study (top left panel). Similarly, we observed a positive relation between having the \textit{urge} to smoke and momentary smoking by the next assessment prior to the quit attempt that diminished during the three weeks following the quit attempt, before sharply increasing during the fourth week post-quit (top right panel). Throughout the assessment period, we observed a positive relation between \textit{negative affect} and momentary smoking by the next assessment that increased during the first week post-quit, leveling off at an odds ratio of 1.75 until the third week after the quit attempt.  We additionally found a positive relation between \textit{cigarette availability} and the odds of momentary smoking by the next assessment that strengthened over the assessment window. For a 1 SD increase in \textit{cigarette availability}, the odds of momentary smoking by the next assessment increased by 300\% for the typical subject one week after the quit attempt, holding all else constant. In the two lower panels of Figure \ref{fig:curves} we observe a relatively weak, oscillating effect of being \textit{bored} and \textit{interacting with a smoker} on momentary smoking by the next assessment, respectively. In addition to these effects, the model identified a constant effect for \textit{alcohol consumption} in the last hour and \textit{motivation to quit smoking} over the assessment period. A similar set of fixed effect relations were identified by our model without the SDP prior, with the exception of not selecting being \textit{bored}. 

Compared to standard TVEMs, our approach deconstructs the structure of the relations between risk factors and smoking behaviors over time, aiding the interpretation of the underlying trends. This information may help the development and evaluation of tailored intervention strategies targeting smoking cessation using mHealth data. For example, \textit{negative effect} has an obvious positive association with momentary smoking by the next assessment that wavers around an odds ratio of 1.2 to 1.5 for a majority of the study. However based on Figure \ref{fig:curves}, it is unclear whether or not the effect linearly diminishes over time. By performing selection on the main effect, linear interaction, and non-linear interaction terms separately, we are able to obtain an actual point estimate for the constant effect of \textit{negative affect} (OR 1.40) as opposed to subjectively assuming a range of values from the plot. Additionally, since the linear interaction term was not selected, we can claim that the effect was not linearly decreasing over time and that it was simply wavering around the constant effect throughout the study.

Tables \ref{tab:one} and \ref{tab:two} present the estimated variances and corresponding 95\% credible intervals (CI) for the random effects selected using SDP priors and traditional spike-and-slab priors, respectively. Using SDP priors, our method identified a random main effect for \textit{urge} to smoke, \textit{cigarette availability}, being \textit{bored}, and \textit{motivation to quit smoking} as well as a random linear interaction between being assigned to the \textit{SmartT} treatment group, \textit{interacting with smokers}, and being \textit{bored} with time. Thus even though we did not discover an overall difference in the odds of momentary smoking by the next assessment for those assigned to the \textit{SmartT} treatment versus \textit{usual care}, we observed evidence that the subjects responded differently to the \textit{SmartT} treatment across the assessment window. With the traditional spike-and-slab priors, we found similar results overall. However, the model only selected a random main effect for \textit{interacting with smokers} and additionally suggested a random effect for \textit{anxiousness}.  

By using SDP priors, our approach is capable of clustering covariates that share similar linear trends with momentary smoking by the next assessment over time. In practice, this information can be used to help construct decision rules when designing future intervention strategies. In our analysis, only five main effect and linear interaction terms were selected, and each of them were allocated to their own cluster. With this knowledge, researchers can prioritize targeting risk factors based on their relative strength of association at a given moment. Had some of these risk factors' effects been clustered together, researchers may rely more heavily on other pieces of information, such as the cost or success rates for a particular intervention strategy, when assessing which risk factors to target during a high-risk moment.

\begin{table}
\centering
\footnotesize
	\begin{tabular}{ccc}
		\hline
		\textbf{Random Effect}       & \textbf{$\hat{\sigma}^2$} & \textbf{95\% CI} \\ \hline
		Intercept                    & 0.923                       & (0.539, 1.528)     \\
		Urge                          & 0.152                       & (0.031, 0.278)     \\
		Cigarette Availability  & 0.865                      & (0.394, 1.467)     \\
		Bored                         & 0.183                      & (0.076, 0.398)     \\
		Motivation to Quit Smoking    & 0.156                      & (0.045, 0.311)     \\
		SmartT $\times$ Time            & 0.077                      & (0.010, 0.210)     \\
		Interacting with a Smoker $\times$ Time & 0.016                      & (0.002, 0.050)     \\
		Bored $\times$ Time              & 0.002                      & (0.000, 0.005)     \\ \hline
	\end{tabular}
	\caption{Smoking Cessation Study: Estimated variances with corresponding 95$\%$ credible intervals (CI) for selected random effects with SDP priors based on MPPI $\geq 0.50$.}
	\label{tab:one}
\end{table}

\begin{table}
\centering
\footnotesize
	\begin{tabular}{ccc}
		\hline
		\textbf{Random Effect}       & \textbf{$\hat{\sigma}^2$} & \textbf{95\% CI} \\ \hline
		Intercept                    & 1.317                       & (0.676, 2.487)     \\
		Urge                   & 0.099                      & (0.011, 0.248)     \\
		Cigarette Availability                  & 0.905                     & (0.503, 1.607)     \\
	    Interacting with a Smoker                  & 0.848                    & (0.286, 1.924)     \\
		Bored                  & 0.244                      & (0.065, 0.517)     \\
		Anxiousness                  & 0.140                    & (0.001, 0.361)     \\
		Motivation to Quit Smoking                & 0.212                      & (0.076, 0.448)     \\
		SmartT $\times$ Time    & 0.062                      & (0.016, 0.155)     \\
		Bored $\times$ Time & 0.002                      & (0.000, 0.004)     \\ \hline
	\end{tabular}
	\caption{Smoking Cessation Study: Estimated variances with corresponding 95$\%$ credible intervals (CI) for selected random effects with traditional spike-and-slab priors based on MPPI $\geq 0.50$.}
	\label{tab:two}
\end{table}

Similar to previous studies investigating the temporal relation between risk factors and smoking behaviors around a quit attempt,  our results show a convex relation between \textit{urge} to smoke and momentary smoking after the quit attempt, a positive association with \textit{cigarette availability} throughout the quit attempt, and a positive, increasing relation between \textit{negative affect} and momentary smoking during the first week after the quit attempt 
\citep{koslovsky2017time,vasilenko2014time}. Existing TVEMs approaches, however, typically model the repeated measures structure of the data by simply including a random intercept term in the model, neglecting to investigate random main effects or interaction terms. They also do not incorporate variable selection.  Our approach, on the other hand, delivers insights on how relations vary over time as well as how they vary across individuals. % Our method with and without DP priors took around 7 and 3 hours, respectively.  

\subsection{Sensitivity Analysis}
\label{sec:sens}

To investigate our model's sensitivity to prior specification, we set each of the hyperparameters to default values and then evaluated the effect of manipulating each term on the results obtained in section \ref{sec:case}. For the default parameterization, we set the hyperparameters for the prior inclusion indicators $\boldsymbol{\nu}$ and $\boldsymbol{\lambda}$ to $a_{\nu_t}=b_{\nu_t}=a_{\lambda_d}=b_{\lambda_d} = 1$. For interpretation, $a_{\nu_t}=b_{\nu_t}=1$ implies that the prior probability of inclusion for a fixed effect is $a_{\nu_t}/(a_{\nu_t} + b_{\nu_t})=0.50$. The default values for the variance of the normal distribution for the slab of $\boldsymbol{\beta}_0$ and $\boldsymbol{\beta}^{\circ}$ as well as the base distribution for $\boldsymbol{\beta}^{*}$ were each fixed at $5$. Additionally, the mean and variance for the random effect terms' proposal and prior distributions were set to $0$ and $5$, respectively. The hyperparameters for the concentration parameters $\vartheta$ and $\mathcal{A}$ were set to $a_{\vartheta} = b_{\vartheta} = a_{\mathcal{A}} = b_{\mathcal{A}} = 1$. Lastly, we assumed ${\Gamma} \sim N( \boldsymbol{\gamma}_0 = \boldsymbol{0},\boldsymbol{V}_{\gamma}=\boldsymbol{I} )$. We ran our MCMC algorithm for 10,000 iterations, treating the first 5,000 iterations as burn-in and thinning to every $10^{th}$ iteration for the SDP model, similar to our case study. For each of the fixed and random effects, inclusion in the model was determined using the median model approach \citep{barbieri2004optimal}. 

Since the true model is never known in practice, we evaluated each model parameterization in terms of sparsity levels and overlap with the results reported in the case study section. Specifically, we present the total number terms selected for both fixed and random effects (\# Fixed and \# Random). We also provide the proportion of active risk factors in our case study that were also included by each model and the proportion of inactive risk factors that were also excluded by each model, for fixed (f-IN and f-EX) and random effects (r-IN and r-EX) as well as overall (IN and EX). Results of the sensitivity analysis are reported in Table \ref{tab:four}.  Compared to the results presented in the case study, we found relatively consistent overlap in the risk factors included and excluded by each model overall. We observed moderate sensitivity to hyperparameter values in terms of percent overlap for fixed and random effects of risk factors included in the model, an artifact of the relatively weak associations identified for some of the risk factors. Notably, risk factors showing stronger associations with momentary smoking at the next assessment (e.g., \textit{negative affect}, \textit{cigarette availability}, and \textit{motivation to quit smoking}) were selected by the model regardless of prior specification. Likewise, weaker relations between momentary smoking at the next assessment and risk factors, such as \textit{being bored} and \textit{interacting with a smoker}, were more sensitive to hyperparameters. We also observed that the number of selected fixed and random effects increased (decreased) as the prior probability of inclusion increased (decreased), as expected. In practice, there are a variety of factors researchers should consider when setting the prior probability of inclusion, including the aim of the research study, the desired sparsity of the model, prior knowledge of covariates inclusion, as well as results from simulation and sensitivity analyses to name a few. From a clinical perspective, $\tau^2=10$ reflects a relatively diffuse prior for a given risk factor (i.e., odds ratio between 0.002 and roughly 500). To further investigate the model's sensitivity to regression coefficients' variances, we set $\tau^2=v_0=1000$, and found somewhat similar results to the model with $\tau^2 =v_0 = 10$ overall (i.e., IN = 0.8, EX = 0.8). Here, we unexpectedly found non-montonic behavior in the proportion of included and excluded terms as a function of the coefficients' variance, which might also reflect our model's sensitivity to relatively weak associations as previously noted. In theory, the selection of random effects may be sensitive to the order in which the columns of $\boldsymbol{Z}$ are ordered, since the Cholesky decomposition is itself, order dependent \citep{muller2013model}. In our case study, we did not observe any differences regarding which random effects were selected with a random permutation of the $\boldsymbol{Z}$ columns. In section \ref{sec:sens}, we further demonstrate our model's robustness to the ordering of $\boldsymbol{Z }$ on simulated data.

\begin{table}
\centering
\footnotesize
	\begin{tabular}{cccccc}
		\hline
		& $a_{v_t} = a_{\lambda_d} = 1$, $b_{v_t} = b_{\lambda_d} = 9$ &  & $\tau^2 = v_0 = 2$  &  & $a_{\vartheta} = b_{\vartheta} = a_{\mathcal{A}} = b_{\mathcal{A}} = 0.1$ \\ \cline{2-2} \cline{4-4} \cline{6-6} 
		\# Fixed  &     4                                                        &  &   8                  &  &    6                                                                      \\
		\# Random  &       5                                          &  &    5           &  &      7                                              \\
			IN & 0.60                                                     &  &   0.70 &  &  0.80                                                                             \\
			f-IN & 0.44                                                          &  &    0.67    &  &  0.56                                                                  \\
		r-IN & 0.50                                                          &  &    0.50    &  &  0.83             \\
	 EX  & 0.80                                                &  &    0.60     &  &   1.00                                                                         \\
		f-EX  & 1.00                                                           &  &    1.00     &  &   1.00                                                               \\
 		r-EX   &  0.78                                                  &  &   0.78                 &  &  0.89                                                                       \\
		& $a_{v_t} = a_{\lambda_d} = 9$, $b_{v_t} = b_{\lambda_d} = 1$ &  & $\tau^2 = v_0 = 10$ &  & $a_{\vartheta} = b_{\vartheta} = a_{\mathcal{A}} = b_{\mathcal{A}} = 10$  \\ \cline{2-2} \cline{4-4} \cline{6-6} 
		\# Fixed  &     10                                                         &  &   7           &  &     6                                                                      \\
		\# Random  &       8                                                       &  &    5              &  &      4                                                \\
		 IN & 1.00                                                 &  &    0.60    &  &  0.80                                                                                   \\
		f-IN & 0.78                                                          &  &  0.67                   &  &    0.56                                                       \\
		r-IN  & 0.83                                                    &  &    0.50     &  &  0.33                                     \\
	 EX  & 0.60                                                    &  &    1.00     &  &   1.00                                                      \\
		f-EX   & 0.83                                                         &  &   0.80        &  &    1.00                                                                   \\
		r-EX    &  0.67                                                            &  &   0.78     &  &  0.78                                                                  \\
 \hline
	\end{tabular}
	\caption{Case Study Data: Sensitivity results for the proposed model with SDP across various prior specifications. Total number of terms selected for both fixed and random effects are indicted as \# Fixed and \# Random, respectively. The proportion of active (inactive) risk factors presented in the case study that were also included (excluded) by each model is reported as 
f-IN and r-IN (f-EX and r-EX), for fixed and random effects, respectively. Finally, the overall proportion of active (inactive) risk factors presented in the case study that were also included (excluded) by each model is represented as IN (EX). }
		\label{tab:four}
\end{table}

\section{Simulation Study}
\label{sec:simul}
In this section, we evaluate our model in terms of variable selection and clustering performance on simulated data similar in structure to our case study data. We compared our method with and without SDP priors on varying-coefficient and random effects to two other Bayesian methods which are designed to handle this class of models. The first is the method of \cite{scheipl2012spike}, which has previously shown promising results performing function selection in structural additive regression models using continuous spike-and-slab priors. Their approach differs from ours in that they assume parameter-expanded normal-mixture-of-inverse-gamma (peNMIG) distribution priors for selection, inspired by \cite{ishwaran2005spike}, and design a Metropolis-Hastings with penalized iteratively weighted least-squares algorithm for updating regression coefficients within the logistic framework. A popular alternative to spike-and-slab priors to induce sparsity in high-dimensional regression settings is to assume global-local shrinkage priors on the regression coefficients (see \cite{van2019shrinkage, bhadra2019lasso} for detailed reviews). At the request of a reviewer, we additionally compared our proposed model to a reparameterized version with shrinkage priors \citep{carvalho2009handling}. To achieve this, we replaced the spike-and-slab priors on $\boldsymbol{\beta}$ with horseshoe priors, which belong to the class of global-local scale mixtures of normal priors \citep{polson2010shrink}. For random effects, $\boldsymbol{K}$, we assumed a similar global-local structure for the scale parameters of the folded-normal distribution, $v_0$. To our knowledge, the theoretical properties and selection performance of global-local scale mixtures of non-normal priors have yet to be explored. However we conjectured that the global-local framework should effectively shrink inactive random effects towards zero and allow active terms to be freely estimated. Details of the resulting model and accompanying MCMC algorithm are found in the Supplementary Material. 

We simulated $N = 100$ subjects with $20$-$40$ observations randomly spaced across an assessment window with $t_{ij} \in [0,1]$, without loss of generality. For each observation, we generated a set of 15 covariates, $\boldsymbol{x}_i$, comprised of an intercept term and 14 continuous covariates simulated from a $N_{14}(\boldsymbol{0},\Sigma)$, where $\Sigma_{st} = w^{|s-t|}$ and $w = 0.3$.  To simulate time-varying covariate trajectories, we randomly jittered half of the elements within $\boldsymbol{x}_i$ by $N(0,1)$. Additionally, we set $\boldsymbol{z}_{ij} = \boldsymbol{x}_{ij}$. Thus, each full model contained 15 main effects, linear interactions, non-linear interactions, and random main effects, corresponding to 60 potential terms (or groups of terms for the non-linear interaction components) to select. The first 5 functional terms in the true model were defined as 
\begin{itemize}
	\item $f_1(t_{ij}) = \pi\sin(3\pi t_{ij}) + 1.4t_{ij} - 1.6$
	\item $f_2(t_{ij}) = \pi\cos(2\pi t_{ij}) + 1.6$
	\item $f_3(t_{ij}) = - \pi t\sin(5\pi t_{ij}) + 1.7t_{ij} - 1.5$
	\item $f_4(t_{ij}) =  - 1.5t_{ij} + 1.6$
	\item $f_5(t_{ij}) =  - 1.6,$
\end{itemize}
and the random effects $\boldsymbol{a}_i \sim N(\boldsymbol{0}, \Sigma_{\alpha})$ with  $\sigma_{kk} = 0.75$ and $\sigma_{jk} = 0.4$ for $j,k = 1,\dots, 5$. Thus in the true model, $\psi_{ij} = \sum_{p = 1}^5f_p(t_{ij})x_{ijp} + \boldsymbol{z}_{ij}^{\prime}\boldsymbol{a}_i$. Note that to impose an inherent clustering for the main effects and linear interaction terms, their values were specified to center around $\pm 1.5$. 

We ran each of the MCMC algorithms on 50 replicated data sets, using 7,500 iterations, treating the first 3,750 iterations as burn-in and thinning to every $10^{th}$ iteration for each model. The spline functions were generated similar to our application.  We set the hyperparameters for the inclusion indicators, $a_{\nu_t} =b_{\nu_t} =a_{\nu_t} =b_{\nu_t} = 1 $, imposing a non-informative prior for selection of fixed and random effect terms. Additionally, we fixed the regression coefficient hyperparameters to $\tau^2 = 2$ and $m_0 = 0$ with $v_0 = 10$. For the concentration parameters $\vartheta$ and $\mathcal{A}$, we assumed $a_{\vartheta} = b_{\vartheta} = a_{\mathcal{A}} = b_{\mathcal{A}} = 1$.  Before analysis, the covariates were standardized to mean $0$ and variance $1$.% For the rPQL method, we used smoothly clipped absolute deviation penalties for both fixed and random effects and adopted the same procedure as described in the case study to choose an appropriate value for the tuning parameters. 

For each of the models with spike-and-slab priors, inclusion in the model for both fixed and random effects was determined using the median model approach \citep{barbieri2004optimal}. For the horseshoe model, fixed effects were considered active if their corresponding 95\% credible interval did not contain zero, similar to \cite{bhadra2019lasso}. The 95\% credible interval for random effects  will almost surely not contain zero. As a naive alternative, we assumed a random effect was active in the model if its posterior mean exceeded a given threshold. For the sake of demonstration, we evaluated the performance of the model over a grid of potential threshold values, and presented the results for the best performing model overall. Notably, this solution is only feasible when the true answer is known, which is never the case in practice. Variable selection performance was evaluated via sensitivity (SENS), specificity (SPEC), and Matthew's correlation coefficient (MCC) for fixed and random effects separately. These metrics are defined as 
$$SENS = \frac{TP}{FN + TP}$$
$$SPEC = \frac{TN}{FP + TN}$$
$$MCC = \frac{TP \times TN - FP \times FN}{\sqrt{(TP + FP)(TP + FN)(TN + FP)(TN + FN)}},$$
where $TN$, $TP$, $FN$, and $FP$ represent the true negatives, true positives, false negatives, and false positives, respectively. For the SDP models, clusters of regression coefficients were determined using sequentially-allocated latent structure optimization to minimize the lower bound of the variation of information loss \citep{wade2018bayesian,dahl2017}. Once clusters were determined, clustering performance was evaluated using the variation of information, a measure of distance between two clusterings ranging from $0$ to $\log R$, where $R$ is the number of items to cluster and lower values imply better clustering \citep{meilua2003comparing}. 
 
\begin{figure}
 	\centering
 	\includegraphics[width=4.5in,height=2.1in]{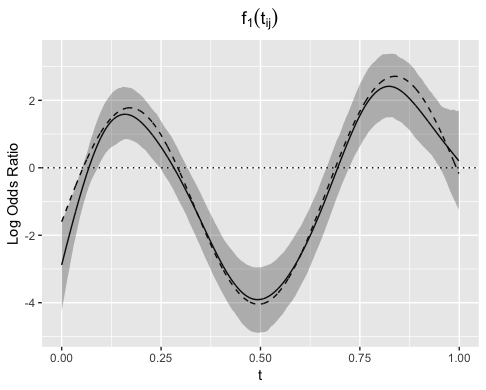} \\
	\includegraphics[width=4.5in,height=2.1in]{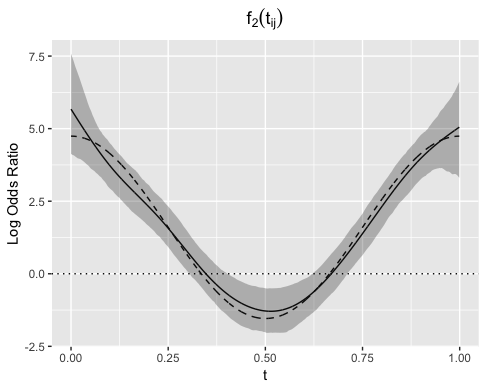} \\
	\includegraphics[width=4.5in,height=2.1in]{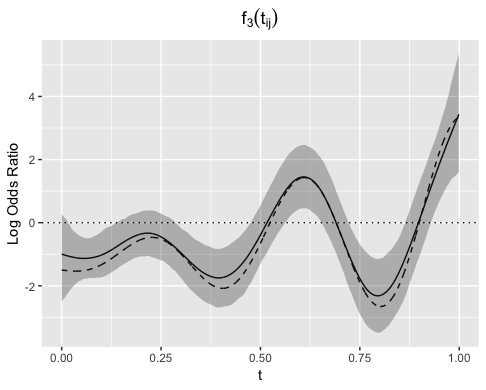} 
	\caption{Simulated Data: Estimated smooth function $f_1(t_{ij}), f_2(t_{ij}), f_3(t_{ij})$ for a randomly selected replicate data set generated in the simulation study. The estimated smooth function is represented by a solid black line with pointwise 95\% credible regions in grey. Dashed lines represent the true log odds ratios as a function of time.
	}
	\label{fig:simul}
 \end{figure}

Figure \ref{fig:simul} presents the estimated smooth functions obtained using our proposed method with SDP priors on a randomly selected replicated data set from the simulation study. Here, $f_1(t_{ij})$ represents the global intercept comprised of a main effect, linear interaction, and non-linear interaction term that were forced into the model. Of interest is the ability of the model to properly select the influential components in $f_2(t_{ij})$ and $f_3(t_{ij})$ and additionally capture their structure. Using the method proposed in \cite{dahl2017} to identify latent clusters of fixed main effect and linear interaction terms, our method successfully clustered the linear interaction in $f_1(t_{ij})$ and the main effects in $f_2(t_{ij})$ and $f_4(t_{ij})$, while incorrectly assigning the linear interaction term in $f_3(t_{ij})$ to its own cluster. Additionally, the main effects in $f_1(t_{ij})$, $f_3(t_{ij})$, and $f_5(t_{ij})$ were appropriately clustered together, while the linear interaction term in $f_4(t_{ij})$ was incorrectly assigning to its own cluster. The remaining, uninfluential terms were all allocated to the trivial group. Despite $f_1(t_{ij})$ and $f_3(t_{ij})$ having similar main effect and linear interaction terms, they are dramatically different in terms of their non-linear interaction terms. However by clustering their underlying linear trajectories, our model with SDP priors was able to uncover similarities in their relations with the outcome over time that traditional approaches would fail to discover.
 
Table \ref{tab:simul} reports results for our proposed method with SDP priors (PGBVSDP), our proposed method without SDP priors (PGBVS), peNMIG, and our model with horseshoe priors (PGHS) in terms of average sensitivity, specificity, and MCC for fixed (fSENS, fSPEC, fMCC) and random (rSENS, rSPEC, rMCC) effects across the replicate data sets with standard errors in parentheses. Additionally for the PGBVSDP model, we provide clustering performance results for fixed (fCLUST) and random effects (rCLUST). Since each of the random effects were simulated similarly, clusterings were compared to a single cluster for the non-zero terms. Overall, the methods had relatively similar results for fixed effects, with PGBVS and PGHS performing the best in terms of sensitivity (1.00 and 1.00) and MCC (0.96 and 0.99), respectively. Our method with SDP priors, PGBVSDP, obtained the highest specificity for fixed effects overall. Given that the maximum possible values fCLUST and rCLUST could take on were 3.4 and 2.7, respectively, we found fairly strong clustering performance for both fixed (0.39) and random (0.92) effects with PGBVSDP. We observed more variability in the selection of random effects across models. Random effect selection sensitivity was significantly lower compared to the fixed effects for all of the models. In terms of specificity (1-false positive rate) for random effects, our methods, regardless of prior formulation, dramatically outperformed peNMIG, with PGBVS obtaining the highest specificity overall (0.96). However, PGBVSDP and PGBVS had lower sensitivity with respect to random effects compared to PGHS. While PGHS performed well separating active from inactive random effects, recall that the truth was used to select the optimal selection threshold. The improved performance of PGBVS, PGBVSDP, and PGHS in terms of variable selection was achieved in considerably less computation time compared to peNMIG. Our core method was able to run 7,500 iterations in a fifth of the time compared to peNMIG, accessed via \cite{scheipl2011spikeslabgam}. Using the SDP priors, which requires additional updates for clustering the regression coefficients, we observed a two-fold increase in computation time for PGBVSDP compared to PGBVS. However on average, the PGBVSDP approach still achieved about a $50\%$ reduction in computation time compared to peNMIG. It is important to note that for comparison, all algorithms were run in series, even though the R package spikeSlabGAM \citep{scheipl2011spikeslabgam} provides functionality to run multiple chains in parallel. 

\begin{table}
\centering
\footnotesize
	\begin{tabular}{cccccccc}
		\hline
		& PGBVSDP &  & PGBVS &  & peNMIG &  & PGHS \\ \cline{2-2} \cline{4-4} \cline{6-6} \cline{8-8} 
		fSENS  &  0.96 (0.09)       &  &   1.00 (0.02)    &  & 0.93 (0.11)         &  &  1.00 (0.00)       \\
		fSPEC  &  0.99 (0.02)       &  &  0.98 (0.02)     &  & 0.94 (0.04)         &  & 0.96 (0.01)        \\
		fMCC   &  0.94 (0.08)       &  &  0.96 (0.05)     &  &  0.83 (0.10)      &  &  0.99 (0.02)         \\
		fCLUST &  0.39 (0.21)       &  & -     &  & -           &  &   -   \\
		rSENS  &  0.76 (0.21)       &  &  0.62 (0.25)     &  & 0.46 (0.23)         &  &  0.86 (0.24)       \\
		rSPEC  &  0.88 (0.10)       &  &  0.96 (0.05)     &  &  0.64 (0.16)         &  & 0.90 (0.11)       \\
		rMCC   &  0.63 (0.26)       &  &   0.64 (0.23)    &  & 0.11 (0.33)         &  &  0.76 (0.21)       \\
		rCLUST &  0.92 (0.50)        &  & -     &  & -        &  &   -      \\
		Time (s) &  4658 (271)       &  & 2235 (46)     &  & 10720 (1116)  &  &  3076 (74)  \\
		\cline{1-8} 
	\end{tabular}
\caption{Simulated Data: Results for the proposed model with and without the SDP on regression coefficients compared to peNMIG \citep{scheipl2012spike} and our model with horseshoe priors \citep{carvalho2009handling}. Results are averaged over 50 replicate data sets with standard deviations in parentheses. }	
\label{tab:simul}
\end{table}

\section{Sensitivity Analysis}
\label{sec:sens}

To assess the model's sensitivity to hyperparameter settings, we set each of the hyperparameters to default values and then evaluated the effect of manipulating each term on selection and clustering performance. For the default parameterization, we set the hyperparameters for the prior inclusion indicators $\boldsymbol{\nu}$ and $\boldsymbol{\lambda}$ to $a_{\nu_t}=b_{\nu_t}=a_{\lambda_d}=b_{\lambda_d} = 1$. The default values for the variance of the normal distribution for the slab of $\boldsymbol{\beta}_0$ and $\boldsymbol{\beta}^{\circ}$ as well as the base distribution for $\boldsymbol{\beta}^{*}$ were each fixed at $5$. Additionally, the mean and variance for the random effect terms' proposal and prior distributions were set to $0$ and $5$, respectively. The hyperparameters for the concentration parameters, $\vartheta$ and $\mathcal{A}$ $a_{\vartheta} = b_{\vartheta} = a_{\mathcal{A}} = b_{\mathcal{A}} = 1$. Lastly, we assumed ${\Gamma} \sim N( \boldsymbol{\gamma}_0 = \boldsymbol{0},\boldsymbol{V}_{\gamma}=\boldsymbol{I} )$. We ran our MCMC algorithm on the 50 replicated data sets generated in the simulation study, using 7,500 iterations, treating the first 3,750 iterations as burn-in and thinning to every $10^{th}$ iteration for the SDP model. 

Results of the sensitivity analysis are reported in Table \ref{tab:sens}. As expected, we found that the sensitivity (specificity) increased (decreased) as the prior probability of inclusion for the fixed and random effects increased. The model did not seem sensitive to the variance assumed for the normal and folded normal priors assigned to the fixed and random effect slab distributions, respectively. Similarly, we found comparable results in terms of sensitivity and specificity for different values of the concentration parameters' hyperparameters. In terms of clustering, we saw marginally better variation of information measures with larger concentration parameter hyperparameters. However across simulations runs, we observed relatively high standard errors in terms of the variation of information measures. To assess potential sensitivity to the order of random effects in our simulations, we re-ran the simulation study with a random permutation of the columns of $\boldsymbol{Z}$. Similar to the case study, we found no evidence of sensitivity to random effect ordering with our model as the results were almost identical to those presented in Table 4 with PGBVSDP (rSENS = 0.76 (0.20), rSPEC = 0.87 (0.09), rMCC = 0.62 (0.20), rCLUST = 0.94 (0.42)).  
\begin{table}
\centering
\footnotesize
	\begin{tabular}{cccccc}
		\hline
		& $a_{v_t} = a_{\lambda_d} = 1$, $b_{v_t} = b_{\lambda_d} = 9$ &  & $\tau^2 = v_0 = 2$  &  & $a_{\vartheta} = b_{\vartheta} = a_{\mathcal{A}} = b_{\mathcal{A}} = 0.1$ \\ \cline{2-2} \cline{4-4} \cline{6-6} 
		fSENS  &     0.92 (0.13)                                                         &  &   0.97 (0.08)                  &  &    0.94 (0.12)                                                                       \\
		fSPEC  &       0.99 (0.02)                                                       &  &    0.99 (0.02)                 &  &      0.99 (0.02)                                                                     \\
		fMCC   & 0.93 (0.10)                                                             &  &    0.96 (0.07)                 &  &     0.94 (0.09)                                                                      \\
		fCLUST & 0.45 (0.30)                                                             &  &  0.35 (0.20)                   &  &     0.45 (0.25)                                                                      \\
		rSENS  & 0.50 (0.20)                                                             &  &    0.79 (0.20)                 &  &  0.54 (0.28)                                                                         \\
		rSPEC  & 0.87 (0.09)                                                             &  &    0.88 (0.08)                 &  &    0.85 (0.10)                                                                       \\
		rMCC   &  0.40 (0.26)                                                            &  &    0.66 (0.23)                 &  &  0.41 (0.29)                                                                         \\
		rCLUST & 1.30 (0.36)                                                             &  &    0.91 (0.44)                 &  &  1.25 (0.50)                                                                         \\ 
		& $a_{v_t} = a_{\lambda_d} = 9$, $b_{v_t} = b_{\lambda_d} = 1$ &  & $\tau^2 = v_0 = 10$ &  & $a_{\vartheta} = b_{\vartheta} = a_{\mathcal{A}} = b_{\mathcal{A}} = 10$  \\ \cline{2-2} \cline{4-4} \cline{6-6} 
		fSENS  &  0.99 (0.03)                                                            &  &   0.96 (0.07)                  &  &    0.94 (0.11)                                                                      \\
		fSPEC  &   0.96 (0.03)                                                           &  &    0.99 (0.02)                  &  &    0.99 (0.02)                                                                       \\
		fMCC   &  0.91 (0.06)                                                            &  &       0.95 (0.07)               &  &    0.93 (0.11)                                                                       \\
		fCLUST &  0.40 (0.20)                                                            &  &    0.39 (0.23)               &  &   0.41 (0.24)                                                                        \\
		rSENS  &   0.85 (0.20)                                                           &  &   0.78 (0.20)                  &  &   0.74 (0.23)                                                                        \\
		rSPEC  & 0.84 (0.10)                                                             &  &   0.89 (0.10)                  &  &  0.86 (0.10)                                                                         \\
		rMCC   & 0.66 (0.19)                                                             &  &   0.67 (0.25)                  &  &   0.60 (0.27)                                                                        \\
		rCLUST & 0.84 (0.49)                                                             &  &  0.89 (0.47)                   &  &  0.97 (0.49)                                                                         \\ \hline
	\end{tabular}
	\caption{Simulated Data: Sensitivity results for the proposed model with SDP on regression coefficients. Results are averaged over 50 replicated data sets with standard errors in parentheses.}
		\label{tab:sens}
\end{table}

\section{Conclusions}
\label{sec:remarks}
In this paper, we have investigated intensive longitudinal data,  collected in a novel, smartphone-based smoking cessation study to better understand the relation between potential risk factors and smoking behaviors in the critical moments surrounding a quit attempt, using a semiparametric Bayesian  time-varying effect modeling framework. Unlike standard TVEMs, our approach deconstructs the structure of the relations between risk factors and smoking behaviors over time, which aids in formulating hypotheses regarding dynamic relations between risk factors and smoking in the critical moments around a quit attempt. By performing variable selection on random effects, the approach delivers additional insights on how relations vary over time as well as how they vary across individuals. Furthermore, the use of non- and semiparametric prior constructions allows simultaneous variable selection for fixed and random effects while learning latent clusters of regression coefficients. As such, our model is designed to discover various forms of latent structures within the data without requiring strict model assumptions or burdensome tuning procedures. 
%In addition, our method is equipped to handle larger data sets than existing methods, making it ideal for intervention studies that collect real-time data from EMA as well as sensors. 
Results from our analysis have confirmed previously identified temporal relations between smoking behaviors and \textit{urge} to smoke, \textit{cigarette availability}, and \textit{negative affect}. They have also identified subject-specific heterogeneity in the effects of \textit{urge} to smoke, \textit{cigarette availability}, and \textit{motivation to quit}.  Additionally, we have found that subjects differed in how they responded to the \textit{SmartT} treatment (compared to usual care), \textit{interacting with a smoker}, and being \textit{bored} over time. This has practical relevance as researchers can use this information to design adaptive interventions that prioritize targeting risk factors based on their relative strength of association at a given moment. They also reinforce the importance of designing dynamic intervention strategies that are adaptive to subjects' current risk profiles.  

Throughout this work, we have demonstrated how our method is well-suited to aide the development and evaluation of future JITAI strategies targeting smoking cessation using mHealth data. The existing \textit{SmartT} algorithm delivers treatment based on the presence of six lapse triggers, which are weighted based on their relative importance in predicting risk of lapse \citep{businelle2016ecological}. The results of this study allow for a more dynamic algorithm that takes into account not only the time-varying relationships between psychosocial and environmental variables and smoking lapse, but the different ways in which individuals experience a quit attempt. For example, the results suggest that providing momentary support to cope with \textit{urge} to smoke and \textit{negative affect} may be more useful if delivered in the early stages of a quit attempt, but become less important by week 4 post-quit. However, messages that address \textit{cigarette availability}, \textit{alcohol consumption}, and \textit{motivation to quit smoking} may be a more important focus for the entire quit attempt. Although the findings for this small sample may not be generalizable to larger, more diverse populations, these methods are the next step in developing a personalized smoking risk algorithm that can inform highly specific, individualized treatment to each smoker.

 It is important to note that selection of a risk factor by our proposed method (or any variable selection technique), does not imply clinical significance. Notably, the point-wise credible intervals often contained odds ratios of one and most risk factors were only influential for brief moments throughout the study period. While these results highlight the importance of understanding risk factors' dynamic relations with smoking to design tailored intervention strategies, we recommend using our method for hypothesis generation in practice and conducting confirmatory studies before generalizing results. 

Compliance rates for EMA studies typically range between 70\% and 90\%, with a recommended threshold of 80\% \citep{jones2006continuous}. In our case study, the compliance rate was 84\%. Additionally, 97.3\% of all assessments were completed once initiated, and subjects were unable to skip questions within an assessment. Since subjects were assessed multiple times per day, nonresponse was attributed more to situational context (e.g., driving) than smoking status. Thus for this study, we found the missing completely at random assumption for missing observations justified. However, future studies may consider the development of advanced analytical methods for EMA data sets that can handle different types of missingness assumptions and other potential biases, such as social desirability bias.

In this analysis, we focus on time-varying effects due to their recent popularity in smoking behavior research \cite{tan2012time,shiyko2012using,vasilenko2014time,koslovsky2017time,lanza2013advancing,shiyko2014modeling}. A promising alternative for investigating the complexity of smoking behaviors around a quit attempt is the varying index coefficient model, which allows a covariate's effect to vary as a function of multiple other variables \citep{ma2015varying}. By incorporating variable selection priors, researchers could identify which variables are responsible for modifying a covariate's effect. Oftentimes behavioral researchers are interested in exploring other forms of latent structure, such as clusters of individuals who respond similarly to treatments or have similar risk profiles over time. Taking advantage of the flexibility and efficiency of our approach, future work could extend our core model to address these research questions by recasting it into a mixture modeling framework. In addition, while we have developed our method for binary outcomes due to their prevalence in smoking behavior research studies, our approach is easily adaptable to other data structures found within and outside of smoking behavior research, such as time to event data \citep{sha2006bayesian} and continuous outcomes. While our method borrows information across regression coefficients, we avoided imposing structure among covariates via heredity constraints, which restrict the model space for higher order terms depending on the inclusion status of the lower order terms that comprise them. Researchers interested in extending our approach to accommodate these, and other forms of, hierarchical constraints may adjust the prior probabilities of inclusion  \citep{chipman1996bayesian}. Lastly, while we were hesitant to present variable selection results for PGHS, due to the limited understanding of global-local priors for non-Gaussian distributions, this showed good results in simulations. Furthermore, when applied to the case study data, we obtained promising predictive performance (i.e., $\widehat{\mbox{epld}}_{HS} = -2955.5$) that warrant future investigation of its theoretical properties.

%The R-package PGBVS contains code to perform the methods described in the article. The package also contains functionality for reproducing the data used in the sensitivity and simulation studies and for posterior inference. The R package is located at (Website Redacted) %\url{https://github.com/mkoslovsky/PGBVS}. 

%Recent advances in Bayesian nonparametric research have generated a variety of methods for drawing inference from posterior samples of latent clusters \citep{wade2018bayesian,dahl2017}. Despite these advances, no methods are currently developed for posterior inference using SDP priors to our knowledge. Existing methods rely on constructing pairwise allocation matrices, which are typically formulated by determining the number of times that elements are assigned to the same cluster in a pairwise fashion across MCMC samples. As a result, they fail to simultaneously accommodate latency in selection as well as cluster assignment. While this limitation does not retract from the benefits of relaxing distributions assumptions, it can result in a majority of the elements being assigned to the trivial cluster when corresponding marginal posterior probabilities of inclusion are relatively low. In future work, we aim to develop a more robust approach for determining cluster allocation from posterior samples in these settings. 

%\appendix

%\section{Appendix section}\label{app}

%This is where the appendix goes. 

\section*{Acknowledgements}
Matthew Koslovsky is supported by NSF via the Research Training Group award DMS-1547433. 

\section*{Supplementary Material}         $\mbox{                                     }$   
\\
\noindent \textbf{R-package for PGBVS:} \\
R-package PGBVS contains code to perform the methods described in the article. The package also contains functionality for reproducing the data used in the sensitivity and simulation studies and for posterior inference. The R package is located at \url{https://github.com/mkoslovsky/PGBVS}.  
\vspace{1cm}

 \noindent \textbf{Supplementary Information:} \\
This file contains a description of the full joint distribution of our model with a graphical representation, a detailed description of our proposed MCMC algorithm with and without SDP priors, and derivations for the prior marginal likelihood used to sample latent cluster assignments. Additionally, we include details of the goodness-of-fit analysis for the case study.

\bibliographystyle{imsart-nameyear}
\bibliography{state3}

\end{document}